# Can molecular dynamics be used to simulate biomolecular recognition?

Malin Lüking[1], David van der Spoel[1], Johan Elf[1], Gareth A. Tribello[2]

[1] Department of Cell and Molecular Biology, Uppsala University, Husargatan 3, SE-75124 Uppsala, Sweden

[2] School of Mathematics and Physics, Queen's University Belfast, Belfast, BT7 1NN, UK

**Abstract**

There are many problems in biochemistry that are difficult to study experimentally. Simulation methods are appealing due to direct availability of atomic coordinates as a function of time. However, direct molecular simulations are challenged by the size of systems and the time scales needed to describe relevant motions. In theory, enhanced sampling algorithms can help to overcome some of the limitations of molecular simulations. Here, we discuss a problem in biochemistry that offers a significant challenge for enhanced sampling methods and that could, therefore, serve as a benchmark for comparing approaches that use machine learning to find suitable collective variables. More in particular, we study the transitions LacI undergoes upon moving between being non-specifically and specifically bound to DNA. It is found that many degrees of freedom change during this transition and that the transition does not occur reversibly in simulations if only a subset of these degrees of freedom are biased. We also explain why this problem is so important to biologists and the transformative impact that a simulation of it would have on the understanding of DNA regulation.

## Introduction

The energy landscape of biological molecules often have multiple low energy states which are separated by large barriers. Interesting phenomena such as protein folding typically involve the molecule transitioning between these low energy states[1]. Simulating these phenomena using molecular dynamics (MD) is thus challenging as the barriers between states prevent the system from exploring all of configuration space during the relatively short timescales that are easily accessible on computers[2]. Typically, then only those configurations in a locally-ergodic region in the vicinity of the input structure are visited.

A number of methods have been put forward to overcome the sampling problem described in the previous paragraph. Furthermore, several of these methods are mature and can be routinely used to extract free energies and kinetics for a wide variety of biological applications[3–10]. Many of these methods are reliant on identifying a small number of collective variables (CVs) that describe the transitions. Having suitable CVs is essential when it comes to understanding the results that have been obtained from simulations that sample large configurational spaces. Furthermore, in many methods the CVs play an active role in ensuring a more complete sampling of configuration space. For example, in some methods the temperature of the collective variables[11] are raised and in others a bias that is a function of the collective variables is added[12,13]. Regardless, the net effect of both these types of simulation is to increase the amount of sampling along the CVs. The sampling along other degrees of freedom, meanwhile, is the same as it would be in an unbiased MD simulation.

It is often difficult to identify suitable CVs for running enhanced sampling calculations. Therefore, as is discussed in this special edition, there is interest in using machine learning algorithms to automate the process by which suitable coordinates for analysing and biasing molecular dynamics simulations are identified[14–17]. However, despite all of this effort none of the new algorithms have really taken off in the way that older methods, such as metadynamics or umbrella sampling, that use a small number of CVs, have.

A possible problem with these new algorithms is that identifying CVs that can be used with the older and simpler methods is not particularly difficult for many of the problems that are actively being studied. As an example consider trying to simulate that quintessential rare event nucleation. It is relatively easy to think of CVs that measure the degree of crystallinity in a sample. One can thus find multiple simulation papers where the number of solid-like atoms is forced to increase using metadynamics or some other rare event method[18–20]. Nucleation can thus be studied using enhanced sampling methods that are well established and machine-learned collective variables are not required.

At the other extreme consider trying to simulate a process such as biomineralisation[21]. There is relatively little theory for describing how biological molecules interact with growing inorganic crystals. For problems like these, even if we could use a machine learning algorithm to enhance the sampling, we have no suitable theoretical framework to make sense of what was observed.

When it comes to identifying a problem that can only be solved using free energy methods that employ some form of machine learning we thus find ourselves with something of a conundrum. If we are to use one of these algorithms, the problem we are endeavouring to solve must be simple enough that we have a body of theory that can be used to analyse any results that we obtain. At the same time, however, the problem must also be not so simple that this body of theory offers us an obvious set of collective coordinates to bias.

This paper describes an interesting problem that we believe satisfies the criteria that have been outlined above. As we will explain in the following pages, the experimental evidence on the binding of proteins to DNA leads one to believe that this system must undergo transitions between certain states that have a well-defined physiological function. There are, however, as we will explain, many degrees of freedom that could be involved in this transition. We hope that by using machine learning researchers can untangle the effects that all these competing degrees of freedom have on the process via which the *lac* repressor binds to its specific DNA operator. We would therefore invite all researchers who have developed free energy methods that employ machine learning to attempt to solve the problem that we will lay out over the following pages. If members of this research community rise to this challenge we believe this will provide a way of benchmarking and comparing these new algorithms. Furthermore, these simulations offer a pathway to resolving longstanding problems about how the *lac* repressor identifies its specific binding sequence.

In the following sections we first provide some background information on the problem. We then discuss results from unbiased simulations of the specifically bound complex. This part is followed by discussion of a set of restrained MD simulations that use information from NMR experiments to extract information on the structure of the non-specifically bound complex. We show that a number of collective variables have very different values in these two sets of simulations and in the NMR structure. However, the transition between the two bound states cannot be driven by adding a metadynamics bias on these variables. We thus conclude that some more sophisticated analysis of the unbiased and NMR-restraint simulations is required. In future, it is hoped that a suitable, low-dimensional descriptor that can be used in an enhanced sampling calculation can be extracted through such analysis.

## Background

### Genetic regulation and the example of the lac operon

The biomolecular system that we discuss in this paper is responsible for genetic regulation. Gene regulatory elements were first described by Jacob and Monod in a paper from 1961[22]. They describe the process *via* which a group of proteins that import and metabolize lactose, a galactoside, to be used as an energy source in *E. coli,* are synthesized only if galactosides are present. To understand how this process works, it is important to understand the flow of information from genes to proteins. In accordance with the central dogma[23], information is transferred from DNA and used to synthesize proteins via a messenger (today known as messengerRNA or mRNA). In bacteria this process often involves a cluster of genes, a so-called operon. An operon is a stretch of DNA that codes for several proteins[24]. Importantly, a particular operon is under the control of an operator and a regulatory protein[22]. The way these agents control gene expression in the case of the *lac* operon is illustrated in figure 1. The top panel of this figure shows how the regulator, or repressor, prevents the expression of the structural genes that encode the messengers for the synthesis of a group of proteins (depicted as X, Y and Z). As you can see the R-repressor, which is coloured in cyan, binds to the operator colored in green and thus prevents the RNA polymerase, which is shown in gray, from transcribing the DNA to mRNA. If a beta-galactoside, shown as a blue circle, binds to the repressor, now colored in magenta, the affinity of the repressor to the operator decreases and it unbinds from the DNA. By unbinding the repressor gives way for the RNA polymerase (RNApol) to transcribe the gene[25]. As shown in the bottom panel of the figure the mRNA produced by RNApol is translated into a protein sequence by the ribosome (for more details see this review[26]).

The specificity of the repressor for its operator or operator sequences is remarkable. As Gilbert and Müller-Hill have shown, the operator is just a particular stretch of DNA[27]. The repressor thus works by binding more strongly to some DNA sequences than others despite these other sequences being in a $10^6$ fold excess[28].

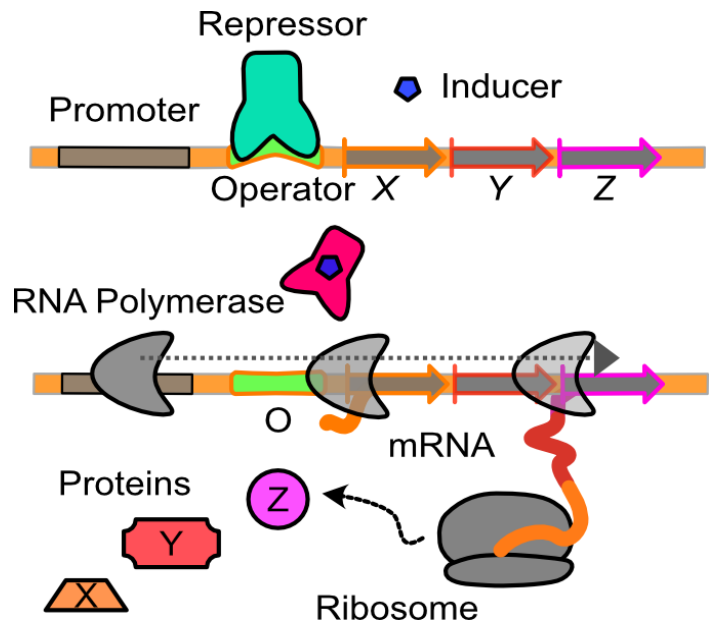

**Figure 1** A simplified representation of the *lac* repressor-operator system showing the induction of the gene expression of structural genes *X*, *Y* and *Z* by an inducer (blue) that binds to the repressor (active-cyan, inactive-magenta). The repressor dissociates and RNA polymerase (RNA pol, gray) transcribes the gene into mRNA, which is then translated into the protein sequence by the ribosome.

<u>Facilitated diffusion</u>

The genetic regulation systems that we have discussed in the previous paragraphs are notably efficient. The kinetics of both repression and induction are both fast with association rates of about 3-5 minutes measured *in vivo* for a strong operator[29]. When measured directly with membrane filter assays the association rate between repressor and operator was found to be about $7 \cdot 10^9$ M$^{-1}$s$^{-1}$. This rapid association rate is remarkable as it beats the theoretical limit for a simple diffusion controlled reaction by an order of magnitude[30]. Some additional process that facilitates the association must therefore be occurring.

Figure 2 illustrates the process of facilitated diffusion that explains these fast binding rates. This figure shows how the freely diffusing repressor quickly binds the negatively charged DNA because of its net-positive charge. The fact that the repressor binds to the DNA non-specifically before it binds the operator site is critical as it reduces the dimensionality of the search and thus explains the rapid kinetics of the process[31–33].

We can imagine facilitated diffusion in the following way: once bound, the repressor moves along the non-specific DNA to its operator site through the various mechanisms that are illustrated in figure 2. These mechanisms include rotation-coupled sliding, where the protein follows the DNA major groove, intersegment transfer, where a protein is transferred, without dissociation, between DNA sites that are spatially close but far apart in sequence space, and hopping, short dissociation and quick reassociation a few base-pairs down- or upstream. All these processes are crucial for the repressor to find the target site quickly and efficiently[34–37]. In fact, it would seem that the processes the repressor performs are balanced in a way that maximises the likelihood of encountering the operator site once the repressor is non-specifically bound to the DNA in the proximity of the operator.

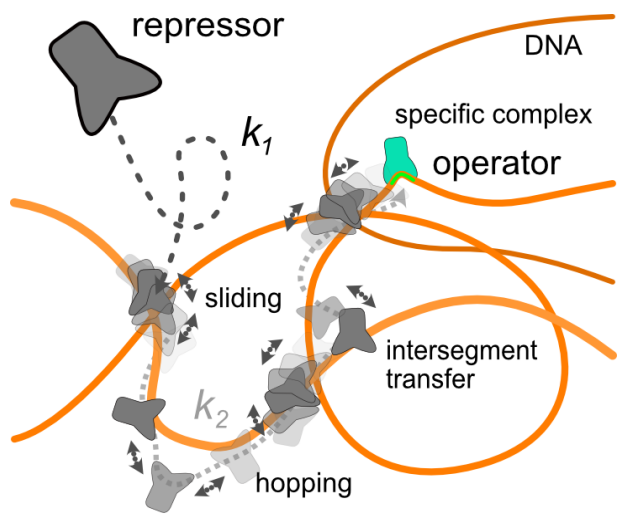

**Figure 2** Free and facilitated diffusion with respective binding rates $k_1$ and $k_2$ that lead to the formation of the specific complex between the repressor (gray → cyan) and the operator (green) after interactions with the non-specific DNA (orange). During facilitated diffusion, the repressor can move along the DNA using different processes: sliding, in close contact with the DNA groves, leading to rotation around DNA, hopping, short dissociation and reassociation, and intersegment transfer between different parts of the DNA.

Given the mechanism for specific binding, illustrated in figure 2, we should be able to rationalize the kinetics of the process by considering the following sequence of two reactions[31]:

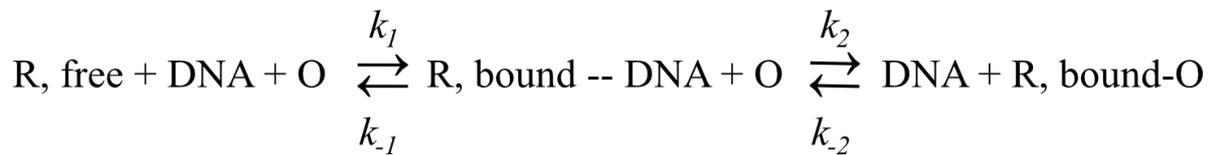

The first step here describes the process by which the repressor binds non specifically to the DNA, described by the forward rate constant, $k_1$, and the backward rate constant, $k_{-1}$. Due to the high amount of DNA though, the protein is nonspecifically bound approximately 90% of the time[38]. The step we focus on in this paper, is the second one in which the repressor transitions in a first order reaction from a nonspecific to specific complex with DNA. As shown in figure 2 and discussed above, this search is carried out using a number of different physical mechanisms. The most important of these mechanisms for our purposes, however, is sliding because the repressor recognises the specific site and consequently transitions into the specific complex while undergoing this process[35,37]. That recognition only takes place while the repressor is sliding is evidenced by the so-called antenna effect, which effectively increases the size of the target to the sliding length[29,31,39].

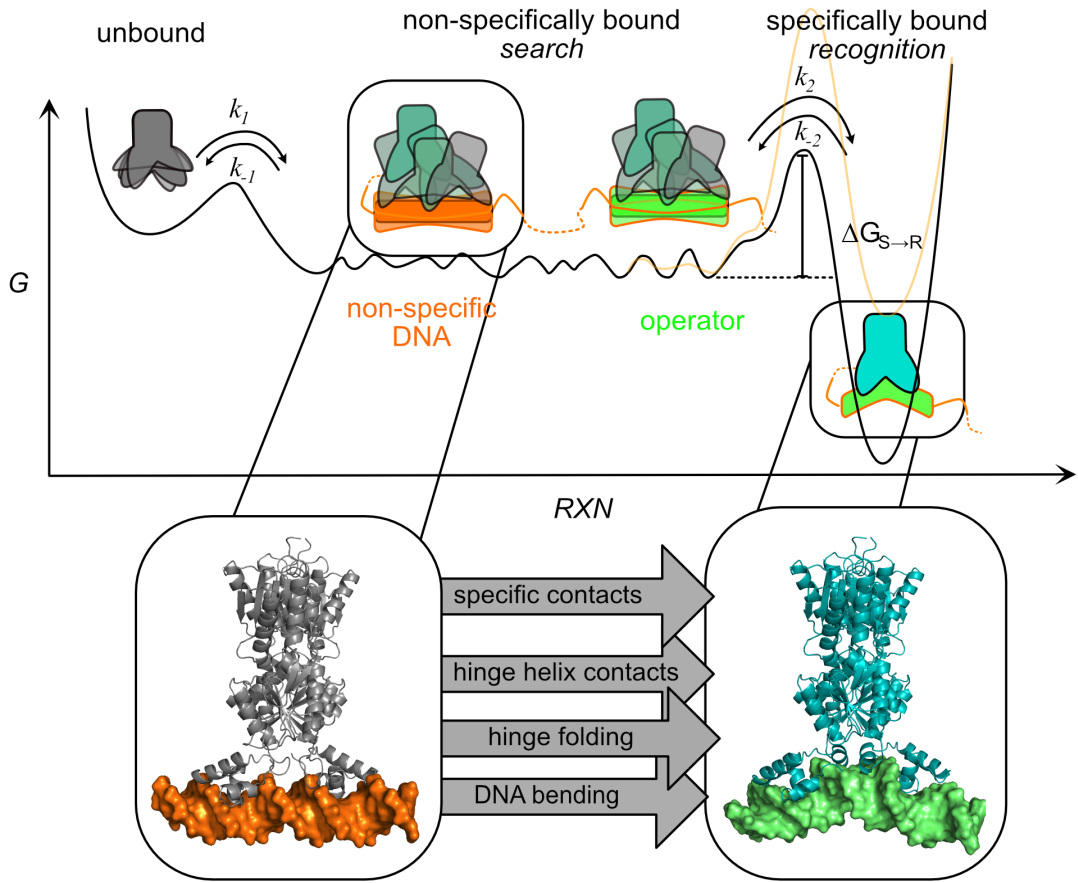

**Figure 3** Free energy along the reaction coordinate of the free to the specifically bound operator. The energies relate to the two-step binding mechanism of the free protein (grey) via the non-specifically bound or *search* state (grey) and finally the specific complex with the protein in the *recognition* state (cyan). The protein binds first to non-specific DNA (orange) and eventually meets the operator (green). At the operator, the transcription factor eventually transitions to the recognition state and the specific complex between the protein and the operator forms. The scheme depicts the barriers and associated transition rates between the three states of the protein. The transparent orange barrier shows a potential barrier for a more weakly bound operator (adapted from ref. [40]). The lower part of the figure depicts which processes could, amongst others, play a role in determining the height of the transition barriers and the associated rate constants that separate the non-specific and the specific complexes.

Figure 3 provides an illustration of the free energy landscape for the reaction mechanism that was described in the last paragraph. The basin on the left of this figure represents the unbound state, which is entropically stabilised because the protein is able to diffuse through a larger volume of configurational space. The middle, rough part of the energy landscape illustrates the non-specifically bound state between the repressor and DNA, which the system enters while the repressor searches for the operator site. This process can be described as a thermally driven random walk in a rough potential[41]. This rough potential is illustrated by the bumpy free energy landscape in the middle part of figure 3. Slutsky and Mirny and others have argued that these bumps must have a height that is less than 1-2 $k_B T$ to achieve fast enough search times when the system is in the non-specifically bound state[42,43].

Slutsky and Mirny have also argued that the energy landscape must be rougher when the repressor is in its specifically bound conformation[42]. This rougher landscape would explain why the repressor ceases its search once the operator is found. The authors argue that the roughness increases upon specific binding because the conformation of the DNA binding protein changes. There are thus two different conformations for the DNA-binding protein: the search state that interacts with non-specific DNA through the electrostatic interaction and the recognition state that binds at the target sequence. When the protein is in the search state it only occasionally forms hydrogen bonds to base edges and thus has a relatively smooth landscape for sliding. When the protein is in the recognition state it tightly binds at the target sequence via hydrogen bonds and van der Waals interactions resulting in a rougher energy landscape for sliding[42,44,45]. As shown on the right of figure 3, this specific protein-DNA complex is characterised by a steep and deep well in the energy landscape[46]. Notice, furthermore, that figure 3 shows a barrier to specific binding. This barrier exists partly because of an entropic cost associated with moving from a non-specifically bound state, where the repressor is free to slide over the full length of the DNA strand, to a specifically-bound, stable state. However, the theory also suggests that there is a transition barrier between the search and the recognition states of the protein. In figure 3, we have thus drawn a barrier between the search and specifically bound states labelled $\Delta G_{S \to R}$[44] to illustrate the energetic cost associated with this conformational change in the protein, the DNA and the surrounding solvent. Also note the additional, higher barrier in the plot in figure 3. The two different barriers stand for two different operators, where a stronger operator supposedly has a lower barrier for specific binding. This relates to experimental studies that have shown that the repressor misses the target site with a certain probability[29] that varies with the operator sequence[40]. It could be argued that a sequence dependent barrier for conformational switching determines the binding probability for a given DNA sequence.

This short review brings us to the core problem that we believe should tantalise those researchers developing machine learning algorithms for sampling; namely, the processes that are responsible for the transition between non-specific and specific binding. Understanding these factors would allow one to determine why conformational change is triggered by specific sequences. To make these determinations it is crucial to know the structure of the non-specifically bound state. It is worth noting that multiple experiments have shown that the energy landscape for sliding is sequence dependent[39,41,47] and that interactions with non-specific sites can affect the search process[48]. This suggests that much can be learned by studying the interaction between one type of repressor and a range of different DNA sequences during non-specific interaction and their transition into specific complexes.

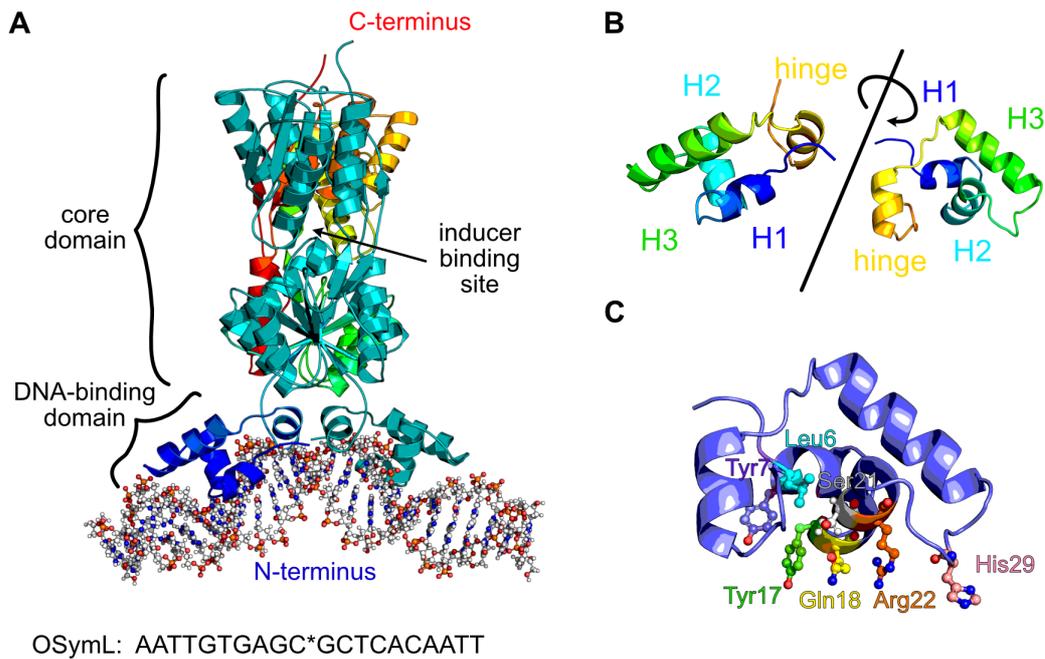

OSymL: AATTGTGAGC*GCTCACAATT

**Figure 4** A) The starting structure for the simulations of the *lac* repressor (LacI) based on the crystal structure with PDB-ID 1EFA[49]. The dimeric repressor is shown in cartoon representation. The first monomer is shown in rainbow colors starting with blue at the N-terminus and going towards red at the C-terminus. The second monomer is colored in cyan. The DNA-binding and core domains as well as the inducer binding site in the center of the core domain are labeled. The DNA is shown in ball and stick representation and the sequence of the OSymL operator is shown below it. The star marks the position that is omitted in this synthetic 20-bp operator when compared to the natural 21-bp operators O1, O2 and O3[50]. B) The monomeric DNA binding domain (DBD) is shown with the labels H1, for helix 1, H2, for helix 2 (the recognition helix), H3, for helix 3 as well as hinge, for the hinge helix/region that will unfold during unbinding. The left panel shows the view onto the DBD that the viewer has on the left DBD in the complex in panel A, the right panel shows a bird view with the angle corresponding to the one in panel C. C) The specifically interacting protein side chains, shown as sticks and spheres, on the DBD that is shown in cartoon representation.

Structural information

The lac repressor that we study in this work is one of the earliest gene regulatory elements known. LacI is natively active as a tetramer but similar sequence recognition has been demonstrated for the dimer[29]. We will thus focus on the dimer that is shown in figure 4A. The dimeric LacI structure shown in this figure consists of a DNA-binding domain (DBD) and a core domain with an inducer binding site that regulates affinity to DNA[51]. Figure 4B zooms in on the DBD, which interacts electrostatically with the phosphate and through hydrogen bonds and van der Waals interactions with the sugars and bases of the DNA[49]. The DBD has a classical motif for a DBP involving 4 helices and 3 turns as well as an N-terminal loop. The first of these helices, Helix 1 (H1) interacts with the DNA through leucine 6 and tyrosine 7[52] (Fig. 4C). Helix 2 (H2), is also called the recognition helix, because it forms most of the specific contacts to the DNA. This helix interacts with DNA through tyrosine 17, glutamine 18, serine 21 and arginine 22[52] (Fig. 4C). Figure 4C also shows the histidine 29 residue that appears in the loop after H2 and can also form specific interactions to the DNA[52,53]. The third helix (H3), is the longest helix in the DBD and is not in direct contact with the DNA. However, this helix carries a number of charged residues that are likely important for the interaction[54,55]. The final and arguably the most interesting helix is the so-called hinge helix, which we investigate further in our simulations.

There are three natural, 21-bp-long operator sites that bind to the LacI dimer: O1, O2 and O3[50]. A range of synthetic operator sites can also be designed, some of which bind more strongly to LacI than the native operator sequences[56]. There are thus multiple crystal and NMR structures for LacI bound to various operator/DNA sequences that can be used as starting points for molecular modeling. When it comes to crystal structures, the strength of binding influences the resolution[57]. The only structure with resolved protein side chains (PDB-ID 1EFA[49]) consequently contains the 20-bp-long sequence OSymL that is known to bind to the protein the strongest. This structure is the starting point for our simulations.

## Methods

<u>General methods: Starting structure, simulation details</u>

We have created a repository on GitHub (https://doi.org/10.5281/zenodo.7627873) to accompany this paper, which contains GROMACS[58,59] input files that can be used to model the systems described in the previous section. The equilibrated structures were prepared based on the crystal structure with the PDB-ID 1EFA[49], that is shown in figure 4.

For crystallization, the structure was stabilized by the anti-inducer ortho nitrophenyl fucoside (ONPF) which binds to the allosteric site in the core domains. Both ONPF molecules were deleted when preparing the structure for simulations. An additional preparation step was performed on the DNA. The crystal structure contains base pairs 4 to 20 of OSymL[49]. For the simulations, base pairs 1 to 3 had to be added manually to form the operator DNA. Two terminal CG base pairs were then added at the ends of the sequence to prevent end effects on the DNA model. The DNA sequence in our simulations is thus 24 base pairs long. For simulations of a complex with a different DNA sequence (NOD), the script basemutator[60] is used to mutate the DNA structure using PyMOL[61].

Histidine 29 (Fig. 4 C) is a critical amino acid residue in the binding between the protein and the DNA[52,53]. There are a number of different protonation states for this residue[62]. We chose to protonate the histidine as PropKa[63] tells us it has a pKa of 7.3 which is very close to the pH of the *E. coli* cell[64]. Despite this, coarse-grained simulations have shown that the protonation of His29 is essential for the sliding motion during the search process[37].

The systems were simulated in the NPT ensemble at 310 K, the optimal growth temperature for *E. coli*[65]. The protein and DNA were solvated in a solution with a 150 mM concentration of KCl and 5 mM concentration of magnesium ions. Magnesium is particularly important because it is known to interact with and stabilize the DNA structure in the cell[66]. The solvation conditions in our simulations are similar to conditions used in earlier simulations of proteins and are representative of the conditions in the *E. coli* cytoplasm[67]. They are also close to the experimental conditions in the first equilibrium binding studies of LacI and the Lac operator carried out by Lin and Riggs[68]. In these experiments the potassium and magnesium concentrations were 180 mM and 3 mM respectively.

All simulations were performed with GROMACS[58,59] version 2019.6 or, in case of metadynamics simulations, with GROMACS version 2021.3. The AMBER99SB-ws force field[69,70] was used to describe the protein interactions and the PARMBSC1[71] force field was used for DNA. Water molecule interactions were simulated using TIP4P/2005[72] and ions were simulated using the parameters from Dang et al.[73]. As discussed in the SI the choice of parameters for the ions is important as standard AMBER force fields overestimate the strength of non-bonded interactions for potassium[74].

The system was prepared in a dodecahedral box. As the LINCS algorithm[58,75] was used to restrain all intramolecular bonds involving hydrogen we could integrate the equations of motion using a 2 fs timestep. Short-range interactions were calculated up to a cut-off of 1 nm. Electrostatic interactions were calculated explicitly below a distance of 1 nm. Beyond that distance they were estimated using the Particle-Mesh-Ewald method[76]. We used the stochastic velocity rescaling thermostat[77] and Berendsen barostat[78] with frictions of 0.1 and 2.0 ps for temperature or pressure coupling, respectively. To equilibrate the system, the protein, DNA and water was initially minimized using the steepest descent integrator. A 500 ps, NVT run at 310 K with positional restraints on the protein and DNA was then performed to equilibrate the solvation spheres. This was followed by a further 500 ps of restrained NPT equilibration at 1 bar during 500 ps. Most restraints on the DNA and protein were removed during production simulations although distance restraints were kept on the terminal base pair hydrogen bonds to prevent fraying of the DNA[79].

**Results and Discussion**

Figure 5 shows the results from 5, 50-ns, unbiased simulations. Figures 5A and 5B show the RMSD for the core domains and the DBDs respectively. These panels illustrate that, in accordance with earlier studies[80], the specific complex is rather stable and the RMSD values are low. This conclusion is further evidenced by figures 5C and 5D which show the distribution of values that were observed for the distance between the hinge helices of the protein and the centre of the DNA and the distribution of values for the number of hydrogen bonds between the protein and the DNA. Both of these distributions are mono-modal, which suggests that only one state in the energy landscape - the specifically bound state - is being explored in our simulation. Much longer simulation times or some form of bias are required to get the complex to transition to the non-specifically bound state.

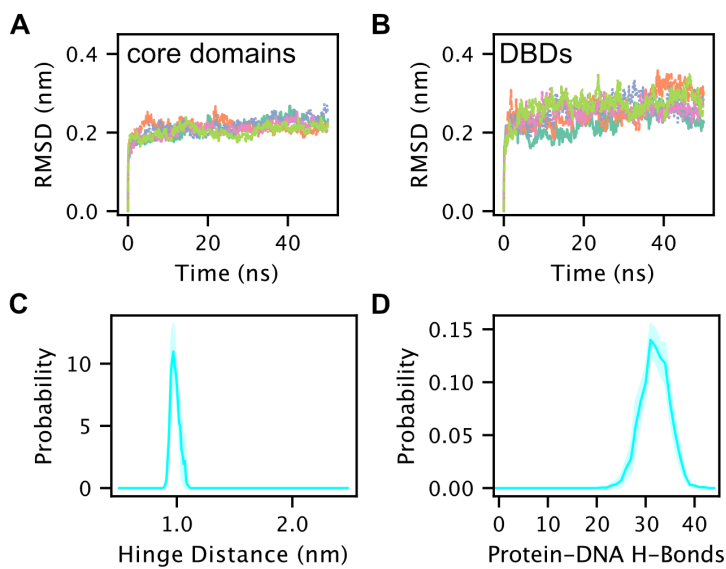

**Figure 5** Results from unbiased MD simulations of the specific complex based on the structure with PDB-ID 1EFA[49]. A) and B) show the RMSD for the core domains and the DBDs. C) and D) show estimates for the histograms together with a shaded area that indicates the 68% confidence interval on our estimate. In C) the histogram is shown as a function of the distance between the hinge center and the center of the DNA. In D) the histogram is shown as a function of the number of protein DNA hydrogen bonds.

## NMR RESTRAINTS

Some structural information on the non-specifically bound complex has been obtained from NMR experiments (PDB-ID 1OSL[81]). These experiments only considered the binding of the two DBDs to the DNA without the core domain. Therefore, valine 52 was mutated to cysteine to stabilise the DBD dimer by a disulfide bridge rather than the interactions between the two core domains.

The NMR experiments provide information on the average distances between particular pairs of atoms or groups of atoms for the non-specifically bound complex rather than a single structure. Multiple structures that have the same average distances between the atoms can then be found. PDB-ID 1OSL thus contains 20 different structures for the non-specifically bound complex, 5 of which are shown in figure 6A. You can see there is substantial overlap between these various different structures for H1, H2 and H3. However, there are considerable differences for the hinge region in the centre of the complex. The large flexibility of the hinge regions is likely a consequence of the removal of the core regions. Furthermore, the structure in this region is likely to be affected by the mutation of valine to cysteine. Figure 6B shows the structure close to the disulfide bridge that holds the two DBDs together in the experiment. Figure 6C then shows what this part of the protein looks like when the wild-type protein is specifically bound to the DNA. You can see that the two DBDs can come closer together when the formation of the dimer is driven by interaction between the core domains and the hydrophobic interactions between the two hinge helices rather than an artificial disulfide bridge.

We performed experiment directed simulations to try to determine the structure of the non-specifically bound state in a way that resolved some of the issues with the NMR experiments described above. Flat-bottom harmonic oscillator potentials on the distances and angles that were identified in the NMR experiments were included as constraints in these simulations (PDB ID 1OSL[81]). To set up these restraints we obtained NMR-STAR files[82] from the protein data bank and then used the method and scripts developed by Sinelnikova and Van der Spoel[83] to construct a restraint file for GROMACS. These restraints force the system away from the initial crystal structure and towards configurations that are supposed to resemble those seen in the experiments. Furthermore, because we remove all the restraints on residue 52, we address the most obvious issues with the mutation that were described in the previous paragraph.

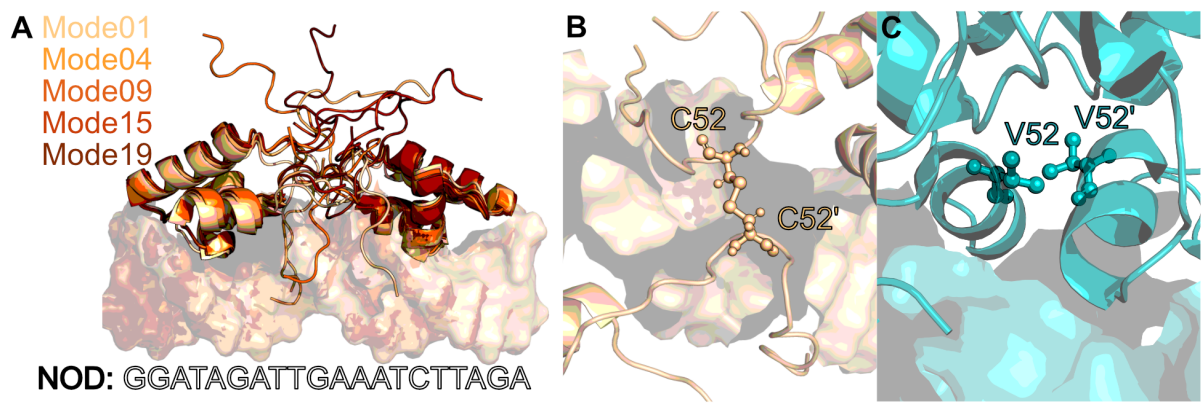

**Figure 6 The NMR structure:** A) 5 of the 20 different modes of the DNA-binding domain in the NMR structure with a non-specific DNA sequence. The monomers are truncated from residue 62 and a cysteine-bridge ensures the dimerization of the transcription factor. B) Bird view and zoom on the central part of panel A. The cysteine bridge between the hinge regions of the DNA-binding domains in the NMR structure is shown for mode 1 using a sticks and spheres representation. C52 and C52' mark C52 in monomer A and B, respectively. C) The interactions between the valine 52 residues in the two monomers in the wild-type protein. V52 and V52' mark V52 in monomer A and B, respectively. In all figures, the protein is shown in cartoon, the DNA in surface representation and individual residues are shown as spheres and sticks. The figures have been produced in PyMOL based on structure with PDB IDs 1OSL[81] and 1EFA[49].

Two separate restrained MD simulations were performed for the specific DNA sequence OSymL, that is known to bind strongly to the DBD of LacI, and NOD, that binds weakly to the DBD [84]. For each sequence 3, 300 ns replicas of restraint simulation with the set-up described in the previous paragraph were performed. A force constant of 1500 kJ mol$^{-1}$ nm$^{-2}$ was used for all NMR restraints and the time constant for averaging the restraint forces was 100 ps. Average properties were then computed from the last 50 ns of all simulations.

RMSD values of 4.7 +/- 0.3 Å and 4.8 +/- 0.2 Å were observed for the atoms in the core domains of the proteins bound to NOD and OSymL, suggesting that the restraints have some effect on the structure in these parts of the protein. Larger RMSD values of 7 +/- 2.6 Å and 7 +/- 0.8 Å are observed for the DBDs, which suggests that the structure in these parts of the protein have changed more. These changes in structure are further quantified in figure 7, which compares (i) the distribution for various structural properties in our restrained simulations, (ii) the distributions of those properties for the specifically bound complex and (iii) the average value of these quantities for the 20 different modes in the PDB file that contains the NMR structure of the non-specific complex[81]. The values these structural quantities take in the crystal structure of the specifically bound complex and the values they take for the NMR structure are also listed in table 1. Table 1 clearly shows that the values of all these structural quantities change when the protein changes from being specifically bound to being non-specifically bound. By contrast, figure 7 shows that some, but not all of these structural quantities, change when the NMR restraints are active and that some variables undergo drastic changes in values that are not seen in the experiments. The picture of unbinding that emerges from our restrained simulations is therefore incomplete.

**Table 1** The values the collective variables that we have investigated in this work take for the experimental NMR structure of the non-specifically bound state and the experimental crystal structure for the specifically bound state. There are multiple configurations in the PDB file that describe the NMR structure so we calculated the CVs for all the reported structures and reported the mean and standard deviation of these values.

| Collective Variable | Non-specific Complex (NMR) | Specific Complex (crystal) |
|---|---|---|
| Number of Specific Contacts | $12.6 \pm 0.58$ | 14 |
| Number of Hinge Helix Contacts | $1.9 \pm 0.75$ | 7 |
| No. of Alpha Helices in Hinge Region | $0.1 \pm 0.1$ | 4 |
| DNA Bent (°) | $1.1 \pm 0.4$ | 41 |
| Hinge Helix to DNA Distance (nm) | $2.1 \pm 0.5$ | 1 |
| Protein-DNA Hydrogen Bonds | $27 \pm 5$ | 33 |

We can better understand the incomplete picture that emerges from the simulations with the NMR restraints by working through each panel of figure 7 in turn. Figure 7A shows the distribution for the number of specific contacts between the DNA bases and protein side chains. In figure 7A the top plot is for NOD and the bottom is for OSymL. NMR studies showed that when the specific complex is formed, Leucine 6, Tyrosine 7, Tyrosine 17, Glutamine 18, Serine 21, Arginine 22 and Histidine 29 interact with the DNA[52,81]. The contact map CV that was used to construct figure 7A was calculated using the distances between these residues and the atoms in the DNA bases that are closest to them in the specific complex (see the SI for a more detailed description and table S1 for the contact parameters for different systems). Figure 7A clearly shows that the number of specific contacts decreases substantially when the NMR restraints are active. However, this behaviour is at odds with the behaviour that is seen in the experiments. The experimental structures suggest that many of these contacts are still present when the protein is non-specifically bound and it is unclear whether this is an artefact in the experiment or a real feature of the non-specifically bound state.

Figures 6A and 4A show the structures of the non-specifically bound and the specifically bound complexes, respectively. The largest differences in these structures appear in the central part of the protein. As has been discussed in previous experimental and simulation work[81,85,86], when the protein is non-specifically bound this hinge region is unstructured and flexible. By contrast, when the protein binds, two alpha-helices form that are bound together tightly[49]. Figures 7B and 7C show histograms for collective variables that measure these differences. Figure 7B shows histograms as a function of the number of contacts between the left and right hinge region for the specifically bound state (top and bottom) and the non-specifically bound state with NOD (top) and OSymL (bottom). The CV here measures the number of native contacts involving Arginine 51, Valine 52, Alanine 53, Glutamine 54, Glutamine 55, Leucine 56 and Alanine 57 that are formed. You can clearly see that all these contacts are present when the protein is specifically bound. This result is in agreement with the experimental information in table 1, which shows that all these contacts are formed when the protein is specifically bound. However, table 1 also shows that many of these contacts are broken when the protein is non-specifically bound. This observation is at odds with what we see in our simulations - the majority of these contacts remain intact in the simulations when the NMR restraints are active. The introduction of the NMR restraints is thus doing little to disrupt the structure of the protein in this region.

The conclusion that the NMR restraints do little to disrupt the structure of the protein in the hinge region is further evidenced by figure 7C, which offers an analysis of the secondary structure content for this part of the protein. We measured the helix content using the ALPHARMSD CV that was developed by Pietrucci and Laio[87]. Figure 7C confirms that 4 helices are formed by the hinge region when the protein is specifically bound and that they remain intact when the NMR restraints are active[84,88]. By contrast, in the NMR structure these helices are basically not present.

Figure 4A shows that the DNA is bent in the specifically bound complex. However, the DNA is straight in the NMR structure of the nonspecific complex[89]. Experimental studies observe bending upon specific binding to LacI for all operators[49,84], which suggests that DNA bending is a feature of specific binding. In figure 7D the histogram of DNA bending angles that were visited in our simulations of the specific complex and our simulations that included the NMR restraints are shown. You can clearly see (top panel) that in the simulations with the NMR restraints the bending angle of the DNA is similar to or even larger than the angle that is seen for the specific complex. In other words, the DNA does not straighten.

Figure 7E shows histograms for what is perhaps the most obvious CV for measuring the degree of binding between the protein and the DNA: the distance between the hinge helix and the central base of the DNA. Naively, one would expect this distance to be small when the protein is specifically bound and large when the protein is non-specifically bound as it is for the crystal and NMR structures respectively (see Tab. 1). When the NMR restraints are active this CV does not change. The protein remains trapped in a conformation with bent DNA that has the helices inserted into the minor groove.

The final panel of figure 7 (panel F), shows the effect the NMR restraints have on the number of hydrogen bonds between the protein and the DNA. You can clearly see that there are far fewer hydrogen bonds when the NMR restraints are active. You would expect the number of hydrogen bonds between the protein and the NMR to decrease when the system moves from being specifically bound to non specifically bound but table 1 shows this is not what is seen in the experiments. Consequently, introducing the restraints from the NMR simulations breaks many hydrogen bonds that the experiment tells us are not broken in the non-specifically bound configuration.

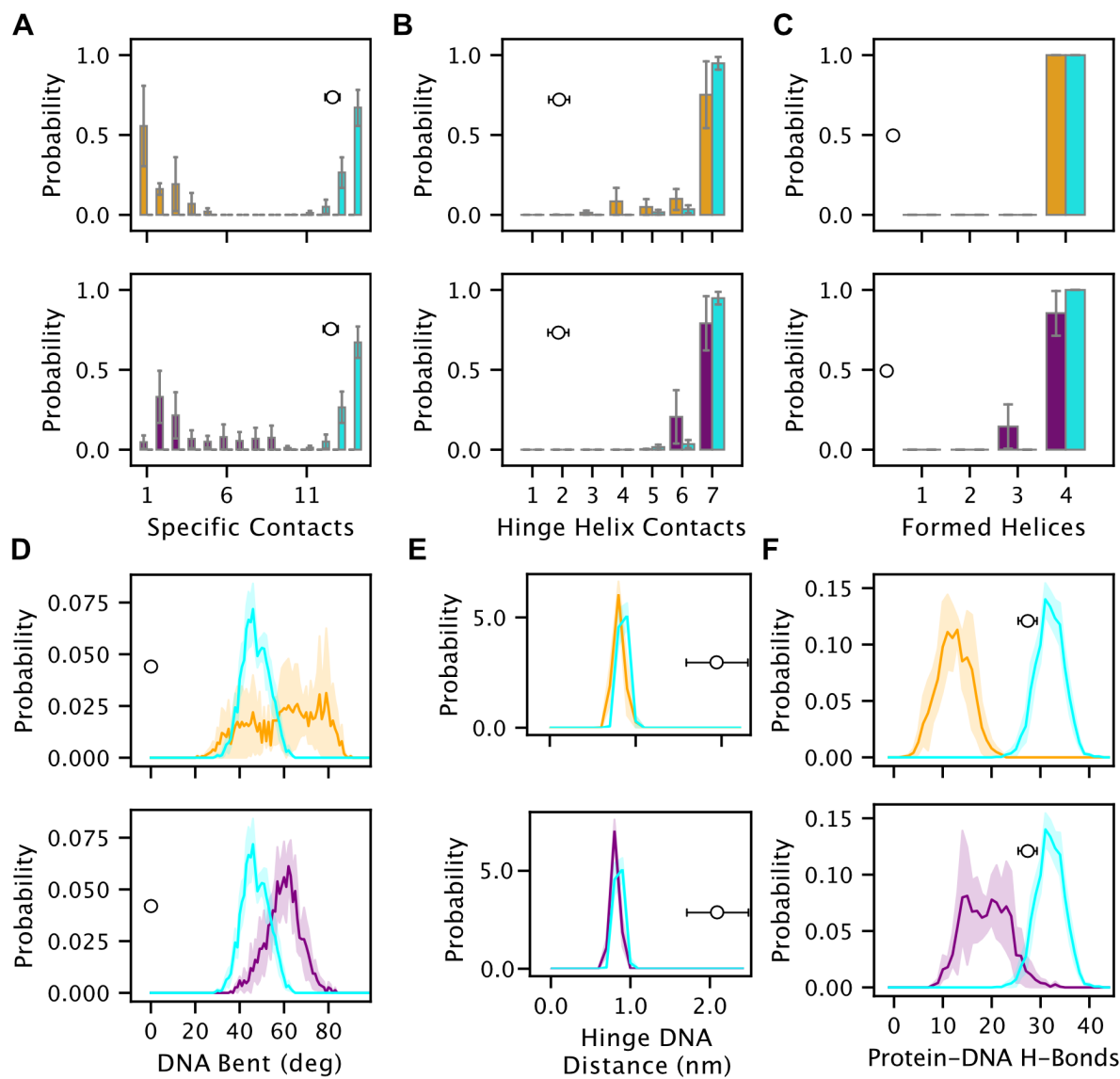

**Figure 7** A comparison between results from MD simulations starting from the specific complex containing the non-specific sequence NOD (orange) and the operator sequence OSymL (purple) with distance restraints based on NMR data and unbiased runs (cyan) of the specific complex with OSymL. Panels A) to F) show the average distributions of different collective variables that could be used to visualize the unbinding process in the metadynamics simulations (see text). The error bars/shaded areas show the standard error of the mean between the 5 replicas. The data used to construct the histograms is taken from unbiased simulations of 50 ns length (see Fig. 5) and the last 50 ns of the NMR

simulations. In all plots the mean and standard deviation for the distribution of CV values in the 20 modes in the NMR structure of the non-specific complex is shown as a circle with whiskers.

To summarise, in the structure that emerges from the simulation with the NMR restraints many of the specific contacts and hydrogen bonds between the protein and the DNA are broken. It would appear that these restraints sever contacts and hydrogen bonds that are present in the the structures of the specifically and non-specifically bound configurations that have been determined from experiment, which is not ideal. To make matters worse, disruptions to the structure of the hinge helices and changes in the degree of DNA bending that are seen in the NMR structure of the non-specifically bound state are not seen in our simulations. When the NMR restraints are active the repressor's hinge helices are left trapped in a tight interaction with the minor groove. If anything, the breaking of the specific contacts and hydrogen bonds between the protein's DNA binding domains and the DNA make the interactions in the centre of the complex even more pronounced. This suggests that the order in which hinge helix unfolding, DNA straightening and contact breaking occur is important. If that is the case, then it is not surprising that the structures for the non-specifically bound complex that emerge from a simulation that was started from the specifically bound state contain features that would not be expected to occur in the non-specifically bound state. The NMR restraints did not drive the system over the barrier between the specifically bound and non-specifically bound state as they serve only to make the non-specifically bound state more thermodynamically favourable. What these restraints do when the system is in the specifically bound state - far from the non-specifically bound state - is unclear. To achieve transitions between the two states some other method is required.

**METADYNAMICS**

We used well-tempered metadynamics[90,91] simulations to try to force the protein to transition from the specifically bound configuration to a non-specifically bound configuration. GROMACS 2021.3[58] patched with PLUMED 2.7.2[92] was used to perform these simulations. In all calculations, the bias factor was set equal to 10 and hills with an initial height of 1.3 kJ mol[-1] were added every 100 MD steps. The width of these Gaussian hills were set using the geometric adaptive framework from Branduardi *et al.*[93] with a sigma parameter of 0.25 nm. The minimum allowed width for the hills was always equal to 0.05 CV units.

The difficulty with metadynamics simulations is to identify a suitable CV for driving the system between the states of interest. As detailed in the following sections, we tried running metadynamics with a number of different CVs but were unable to obtain converged results. The comparison between the distributions of configurations that were visited in the metadynamics simulations with the distributions that were obtained from the unbiased simulations, the simulations that were performed with the NMR restraints and the NMR structure, that is presented in the next few sections, is, nevertheless, useful as it offers further insights into the qualitative behaviours of this system.

## CMAP

Figure 8 shows the results from three simulations in which a metadynamics bias was applied on the number of specific contacts between the protein and the DNA[94]. Histograms as a function of the number of specific contacts are shown in figure 8A. This figure shows that the metadynamics bias does cause the number of contacts to decrease slightly. Figure 8B also shows that the metadynamics bias causes protein-DNA hydrogen bonds to break. Fewer protein-DNA hydrogen bonds break during the metadynamics simulation than in the simulation with the NMR restraints. In this regard the structures visited during the metadynamics simulations are more consistent with the structure of the non-specifically bound state that emerges from the NMR experiments. However, figure 8C illustrates that there are still enormous differences between the ensembles of structures that are emerging from simulations and experiments. This figure shows the distribution for the distance between the hinge helix and the DNA. You can see that the distances that are explored in the metadynamics simulation are very close to the distances that are observed in unbiased simulations and that they are much shorter than the distances that are seen in experiment. Consequently, even though the number of specific contacts decreases, it seems unlikely that this simulation is exploring states that correspond to the non-specifically bound configuration.

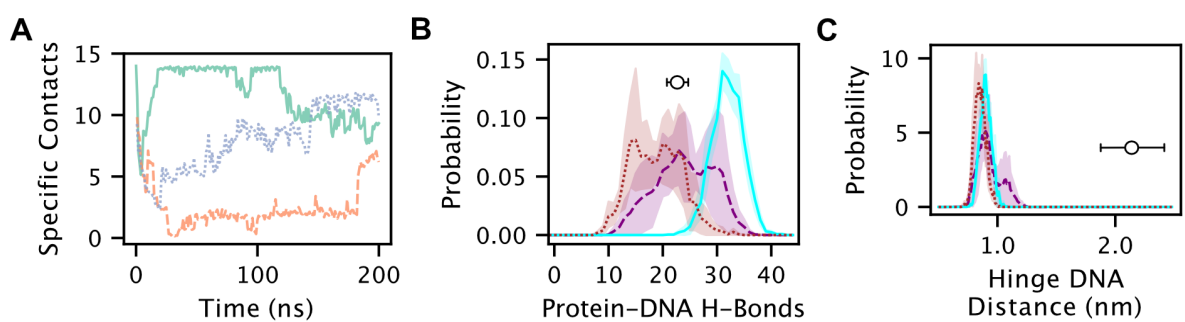

**Figure 8** Results from metadynamics simulations using the number of specific contacts between protein and DNA as a collective variable. A) shows the time series for the number of specific contacts for the three replicas (1-green; 2-orange, dashed; 3-blue, dotted). B) and C) show estimates for the distribution of configurations that are sampled in the three metadynamics simulations (purple, dashed), the unbiased (cyan) and the simulations with the NMR restraints (brown, dotted) with a 68% confidence interval (shaded area) and the averages from the 20 modes in the NMR structure as dot with whiskers. In panel B), the histogram is shown as a function of the number of protein DNA hydrogen bonds that are present. In panel C) the histogram is shown as a function of the distance between the center of the hinge helices and the center of the DNA strand.

## BIASING THE EXTENDED CONTACT MAP

As discussed in the section on the simulations with the NMR restraints and as shown in figures 6B and 6C the NMR experiment shows that the hinge region of the DBD undergoes significant structural changes upon unbinding. We incorporated these structural changes in our metadynamics simulations by biasing a new contact map that measured the number of contacts between the protein and the DNA as well as the number of contacts between the hinge helices of the two DBDs. Results from these simulations are shown in figure 9.

Figures 9A and 9B show that the metadynamics bias causes the number of hinge-helix contacts and the number of specific contacts to decrease. Furthermore, there appear to be correlations between these two groups of contacts. In other words, the hinge helix contacts appear to break at the same time as the DNA-DBD contacts.

Figure 9C shows how this new metadynamics bias is not driving unbinding but that new configurations of the system can be observed. This panel shows the histogram as a function of the distance between the center of the hinge helices and the center of the DNA and should be compared to figure 8C. We once again see here that when the bias acts only on the DNA protein contacts this distance does not change much from the value it has in the specifically bound state but that two additional peaks for shorter and longer distances emerge.

Further evidence of partial unbinding is observed in figure 9D. This figure shows histograms as a function of the number of protein-DNA hydrogen bonds. The inset configurations beside this panel show illustrative configurations from each of the three modes in the distribution that was observed in the metadynamics simulation. You can see that the peak with 38 hydrogen bonds corresponds to the specifically bound complex. In structures that appear in the peak around 26 hydrogen bonds the two hinge helices have begun to drift apart and the DNA has started to straighten. Lastly, in structures that have only about 12 hydrogen bonds we often find that one DBD has drifted away from the DNA.

In short, figure 9 demonstrates that the system is driven further when both the hinge-helix and the specific contacts are biased together. However, we are still clearly missing some critical slow degrees of freedom because we do not see the system returning to the specific complex and the hinge DNA distance does not increase by a great deal.

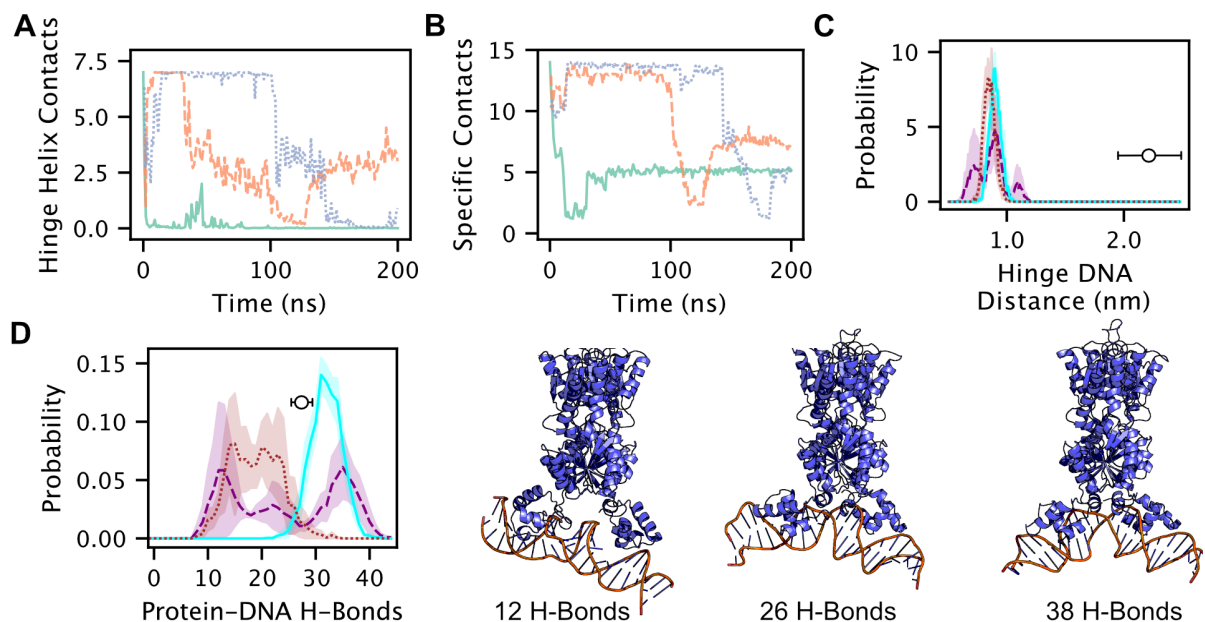

**Figure 9** Results from metadynamics simulations using contacts between the two hinge regions and specific contacts between protein and DNA as a collective variable. A) and B) show the time series of the contacts between the hinge helices and the specific contacts for the three replicas (1-green; 2-orange, dashed; 3-blue, dotted). C) and D) show an estimate for the distribution of configurations that is sampled in the three metadynamics simulation (purple, dashed), the unbiased (cyan), the simulations with the NMR restraints (brown, dotted) with a 68% confidence interval (shaded area) and the averages from the 20 modes in the NMR structure as dot with whiskers. In panel C), the histogram is shown as a function of the distance between the center of the hinge helices and the center of the DNA strand. In panel D) the histogram is shown as a function of the number of protein DNA hydrogen bonds that are present. In D) some representative snapshots of the protein-DNA complex for different numbers of hydrogen bonds are shown.

## BIASSING THE CONTACT MAP AND THE SECONDARY STRUCTURE

The NMR experiments indicate that when the protein is non-specifically bound, the hinge-helices completely unfold so there is no alpha-helical structure in these regions. We found that these helices did not unfold in the metadynamics simulations that were run in the previous section. As shown in the supporting information (Fig. S1), the hinge helices partially unfolded but the ALPHARMSD variable[87] never fell to values of zero. We thus ran a new set of metadynamics simulations in which a metadynamics bias that was a function of both the contact map and the ALPHARMSD was added. The results from this simulation are shown in figure 10.

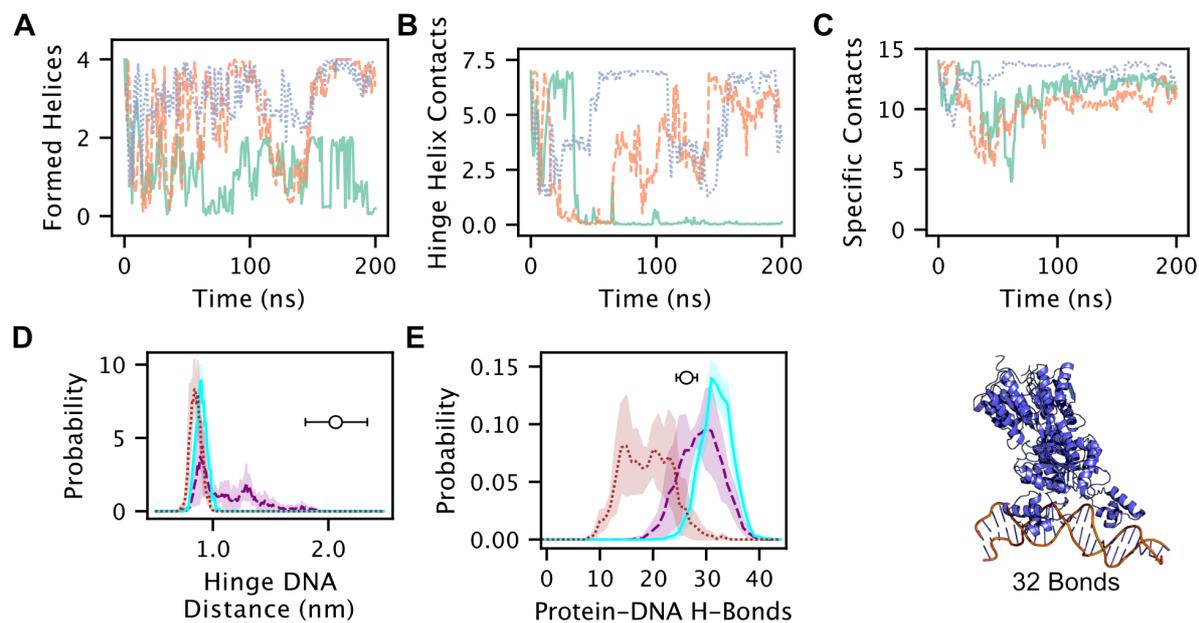

**Figure 10** Results from metadynamics simulations that used the number of specific contacts between protein and DNA and between the two hinge regions and the ALPHARMSD for the hinge regions as collective variables. A), B) and C) show the time series for the number of helices (6 residues) in the hinge region, the number of contacts between the hinge helices and the number of specific contacts between the protein and DNA for the three replicas (1-green; 2-orange, dashed; 3-blue, dotted). D) and E) show estimates for the distribution of configurations that is sampled in the three metadynamics simulation (purple, dashed), the unbiased (cyan), the simulations with the NMR restraints (brown, dotted) with a 68% confidence interval (shaded area) and the averages from the 20 modes in the NMR

structure as dot with whiskers. In panel D) the histogram is shown as a function of the distance between the center of the hinge helices and the center of the DNA strand. In panel E) the histogram is shown as a function of the number of protein DNA hydrogen bonds that are present. In E) a representative snapshot of the protein-DNA complex is shown.

Figure 10A shows that the metadynamics bias is able to drive the hinge helices to unfold and, in some cases, to fold during the simulation. Furthermore, figure 10B shows that the number of contacts between the two hinge helices changes substantially as the hinge helices fold and unfold. However, figure 10C shows that all the activity associated with the folding and unfolding of the hinge helices is not accompanied by any significant change in the degree of binding between the protein and the DNA. The number of specific contacts between the DNA and the protein remains resolutely high, which suggests that the non-specifically bound state is not visited.

This conclusion, that the non-specifically bound state is not visited in these simulations, is confirmed by figures 10D and 10E. Figure 10D shows the histogram as a function of the distance between the center of the hinge helices and the center of the DNA together with the histograms that were obtained from the unbiased and NMR restraint simulation and the average value of this quantity from the NMR experiment. You can see that the histogram we obtain from the metadynamics simulation is similar to the result we get from the unbiased simulations and the NMR simulations. During most of the simulation the distance between the DNA and the protein remains resolutely low.

Figure 10E shows that metadynamics does a better job of determining what happens to the amount of hydrogen bonding between the protein and the DNA than the simulations with the NMR restraints. You can see that most of these hydrogen bonds remain intact during the metadynamics simulations but that this behaviour is also most consistent with the structures that are observed in the NMR experiments. The reason these bonds remain intact becomes clear when you look at the representative snapshot that is shown in figure 10. This figure clearly shows that the bias drives the hinge helices to move away from the DNA and to unfold. However, H1, H2 and H3 remain bound to the DNA strand.

We can thus conclude that unbinding was not observed with this combination of collective variables. It would be easy to argue that this is simply because the simulation was not run for long enough. However, we obtained similar results from longer simulations with similar combinations of CVs. Furthermore, the average value of the bias for the last 100 steps of the simulation is 146 kJ mol$^{-1}$. That such an enormous bias is not able to drive the transition from the specifically bound to the non-specifically bound state suggests that some other degree of freedom needs to be biased in our simulations.

**METADYNAMICS FOR NON SPECIFIC DNA**

The previous three sections showed how difficult it is to design a metadynamics bias that can drive the transitions between the specifically and non-specifically bound states. These sections focussed on illustrative examples, we have tried but not written about other combinations of CVs including the DNA bending angle, various distances between the DNA and the protein, the number of hydrogen bonds, path CVs in contact map space and variously-defined contact maps. Every combination of CVs we tried produced similarly inconclusive results.

We also tried driving the transition between the specifically and non-specifically bound state for a complex where the DNA is known to not bind specifically to the protein, the sequence known as NOD [84]. Figure 11 shows the result from one of these simulations. The metadynamics bias here acts upon the contact map and the ALPHARMSD and is similar to the one that was used to bias the dynamics in the previous section. This bias is necessary as, even though it is not possible to characterize a specifically bound complex with the non-specific sequence experimentally, simulation work has shown that the specifically bound complex with non-specific sequences remains in tact over the course of a $1$-$\mu s$-long simulation[80].

Figure 11C shows that the metadynamics bias is able to drive down the number of specific contacts between the DNA and the protein. There is thus some evidence that the protein begins unbinding from the DNA. Figures 11A and 11B show that this unbinding is accompanied by a loss of structure in the hinge-helix region. As the experiments suggest, the degree of alpha-helical structure in this region decreases as well as the number of contacts between the two hinge helices.

Figure 11D shows the histogram as a function of the distance between the center of the hinge helices and the DNA. You can see that the metadynamics bias drives the system to explore configurations where this distance is large. Furthermore, figure 11E shows that the metadynamics bias also drives the system to explore configurations that have fewer hydrogen bonds between the DNA and the protein. We observed three different types of unbinding in our simulations: either H1, H2 and H3 unbind, the hinge unbinds or the whole DNA binding protein separates from the DNA. In experiments the second of these types, the unbinding of the hinge, is observed.

Even though there are some positives in the results in figure 11 they are still outweighed by the negatives. Recrossing was not observed so we cannot extract relative free energies. Furthermore, any kinetic information that we extract from these simulations is likely spurious as the metadynamics bias is very large and basins are likely to be overfilled. However, there is possibly a deeper problem with this simulation than even the issues with the metadynamics. The fact that specific binding to the NOD sequence is not observed in experiments suggests that the configuration that we started the simulation from is at best metastable. If this is the case, our simulation has been started from a configuration that is far from equilibrium. Without a better sense of the free energy surface it is unclear if the simulation would return to anything resembling the initial, specifically-bound configuration even if a set of collective variables that can drive this transition for the specific sequence had been identified.

Although the barrier between the specifically-bound and non-specifically bound complex is probably lower for NOD, it is better to start out by developing coordinates that can sample the forward and backward transitions for the complex with OSymL or weaker operators such as O1[50] where the specifically-bound complex is seen in experiments[49,95]. Using these same coordinates to drive the complex with O1 between the specifically and non-specifically bound states would allow one to untangle the role kinetics and thermodynamics are playing in the experiment.

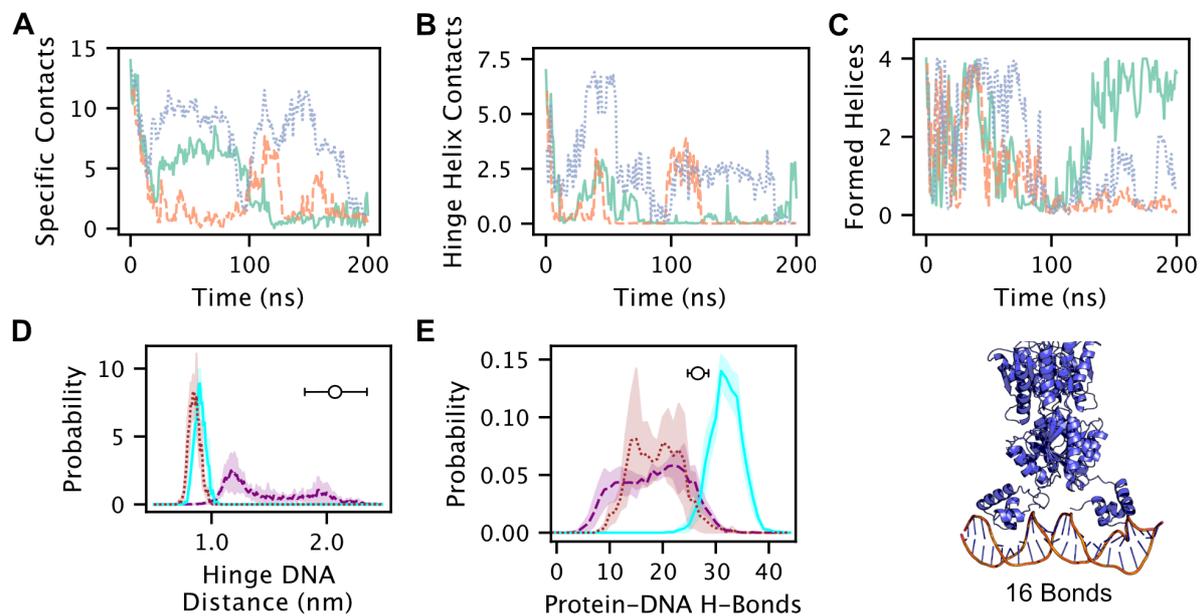

**Figure 11** Results from metadynamics simulations on the complex between LacI and the nonspecific sequence NOD using the number of specific contacts between protein and DNA and between the two hinge regions and the ALPHARMSD for the hinge regions as collective variables. A), B) and C) show the time series of the number of helices (6 residues) in the hinge region, the number of contacts between the hinge helices and the number of specific contacts for the three replicas (1-green; 2-orange, dashed; 3-blue, dotted). D) and E) show an estimate for the distribution of configurations that is sampled in the three metadynamics simulation (purple, dashed), the unbiased (cyan), the simulations with the NMR restraints (brown, dotted) with a 68% confidence interval (shaded area) and the averages from the 20 modes in the NMR structure as dot with whiskers. In panel D) the histogram is shown as a function of the distance between the center of the hinge helices and the center of the DNA strand. In panel E) the histogram is shown as a function of the number of protein DNA hydrogen bonds that are present. In E) a representative snapshot of the protein-DNA complex is shown.

**Conclusions**

If it were possible to perform simulations of the reversible transition between the non-specifically bound and specifically bound complex for LacI and DNA many outstanding questions about gene regulation could be answered. Such simulations would help us to assign a structure to the non-specifically bound complex, which has been the subject of fierce debate for 60 years. Such simulations could also be used to determine how the relative free energies of the specifically-bound and non-specifically-bound states depend on the DNA sequence. This information would resolve arguments about whether specific binding only occurs because the kinetic barrier to formation of the specifically bound complex is low for binding to operators and high for binding to non-operators. In other words, such simulations would tell us if there is a deep minimum in the free energy landscape for specific binding to non-operator DNA sequences that is not seen in experiments because the barrier to accessing it is inordinately high.

The simulations in the previous sections do not answer any of these questions as the CVs do not capture the essence of the transition. However, the work we have described suggests an approach that can be used to tackle this problem. Suitable variables for biasing could be obtained by analyzing the unbiased MD simulations and by performing simulations that use the NMR restraints to identify an ensemble of structures for the non-specifically bound state. We have shown that it is difficult to come up with a single variable that describes all the differences in these two ensembles of structures. Machine learning or dimensionality reduction algorithms are perhaps better suited to tackling this problem. Furthermore, running enhanced sampling calculations on the variables that emerge from this analysis offers a method to test the efficacy of the machine-learned coordinates.

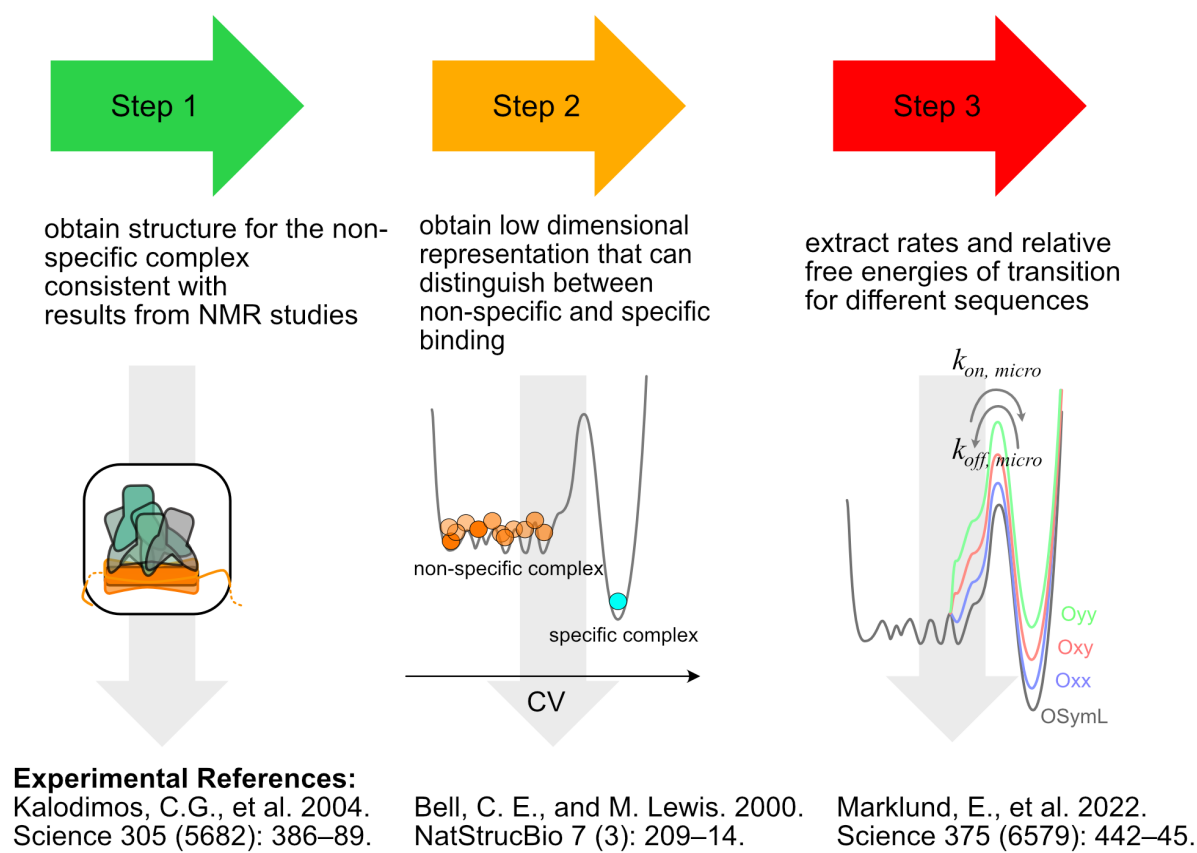

**Figure 12** Schematic illustration and key literature that illustrates the problem that we put forth to the enhanced sampling community. The figure shows the three steps to solving the problem of understanding DNA binding and recognition. The first of these steps is to identify an ensemble of structures for the non-specifically bound state that is consistent with the experimental data from NMR. The second step is to find a CV that is able to drive the system between the non-specifically bound and specifically bound states. The final step is then to calculate the free energy change for this transition and the kinetic rates of unbinding. The rates and relative free energies that emerge from these simulations can then be compared with and used to interpret the ample data that is available from experiments.

A need for benchmark, enhanced-sampling problems that are more complicated than sampling transitions in alanine dipeptide or Lennard Jones 7[96] comes up whenever there is a discussion session at a conference on enhanced sampling. In these sessions, it is argued that more complicated problems are required to test the next generation of enhanced sampling techniques. We believe that the transition from the non-specifically-bound to the specifically-bound protein-DNA complex that we have described in this paper would serve as an excellent challenge to this community. Many degrees of freedom change during the transition and need to be biased for the transition to be observed on an accessible timescale. Furthermore, because the protein and DNA shouldn't unfold during the simulation, it may be difficult to use parallel tempering[97] to enhance sampling of unbiased degrees of freedom. Such simulations would certainly require considerably larger computational resources than we had available for our work.

The 200 ns simulations that were described in the sections on metadynamics took approximately 100 hours on 32 CPU cores. It is thus not prohibitively expensive for those who are interested in tackling this sampling challenge to reproduce the calculations we have performed. We provide example input files on GitHub (https://doi.org/10.5281/zenodo.7794684) to help interested researchers get their simulations started quickly.

Figure 12 shows a graphical summary of the challenge that we propose to the community and highlights the key experimental papers. As the figure shows there are three parts to the problem that we think can be tackled separately. The first problem is to identify the structure of the non-specifically bound complex. In this paper we endeavoured to identify this structure by running simulations with restraints that were informed by the results from NMR experiments and obtained results that were inconclusive. We have provided data on the ensemble of structures that was explored in these simulations but other researchers may find that they can obtain better structures by performing experiment directed simulation in some other manner.

The next step of the challenge in figure 12 is to identify some low dimensional coordinate that is able to distinguish between configurations of the protein that are specifically bound to DNA from those that are non-specifically bound. This coordinate could be a physically-inspired CV like the ones we have discussed in this paper or it could be a machine-learned coordinate that is generated by combining many physically-inspired CVs. Ideally this CV could then be used to drive the protein between the specifically and non-specifically bound states, which, we have shown, is difficult to do using established methods. Such simulations would allow one to extract the relative free energies of the two bound states and the rate constants for the transitions. As the third panel of figure 12 indicates the results from such simulations could be compared with the results from experiments. These simulations may even provide a rationale for better understanding the kinetic rate information that emerges from the experiments. At the very least we hope that if the sampling challenge we have laid out in the preceding sections is accepted widely, there will be lively discussion of the strengths and weaknesses of these various new methods and the mechanism for protein-DNA binding.

## SUPPLEMENTARY MATERIAL

See the supplementary material for details on the characterization of the ion parameters, the preparation of NMR restraint files, a supplementary figure from metadynamics simulations and a detailed description of the collective variables used for biasing and visualisation.

## ACKNOWLEDGEMENTS

We acknowledge support to J.E. from the eSSENCE e-science initiative, the Swedish Research Council (2016.06213), and the Knut and Alice Wallenberg Foundation (2016.0077). The data handling and computations were enabled by resources in Projects SNIC 2021/3-8 (D.V.d.S.), SNIC 2022/3-26 (D.V.d.S.), SNIC 2021/6-268 (J.E.), SNIC 2022/6-261 (J.E.), SNIC 2022/23-373 (M.L.), SNIC 2021/6-294 (D.V.d.S.) and SNIC 2022/6-344 (D.V.d.S.) provided by the Swedish National Infrastructure for Computing (SNIC) at UPPMAX and NSC, partially funded by the Swedish Research Council through Grant Agreement No. 2018-05973. We thank Michele Ceriotti, Lorenzo Agosta and David Wilkins for reading and commenting on early drafts of this manuscript and Martin Zacharias for useful discussions.

## AUTHORS DECLARATION

### Conflict of Interest

The authors declare no conflict of interest.

### Authors Contribution

G.T. conceived the idea of the paper. J.E. conceived the idea of using molecular dynamics simulations to study microscopic binding and unbinding rates for the LacI-operator system. D.v.d.S. conceived the idea of using NMR-restraint to sample the non-specific state. M.L. carried out the simulations, analysis and produced all figures with input from all authors. G.T. and M.L. wrote the manuscript with input from all authors.

## DATA AVAILABILITY

Starting structures, input files for the simulations as well as scripts to prepare NMR restraints and PLUMED input files for different starting structures are openly available at GitHub https://doi.org/10.5281/zenodo.7794684[98]. Trajectories for the different simulations can be found at the SciLifeLab data repository under the following DOI: https://doi.org/10.17044/scilifelab.22348714.v1[99].